\documentclass[aps,showpacs,floatfix,preprintnumbers,nofootinbib,11pt,prd, superscriptaddress]{revtex4-2} 

\usepackage{graphicx}
\usepackage{dcolumn}
\usepackage{bm}
\usepackage{amsfonts}
\usepackage{amsthm}
\usepackage{amsmath}
\usepackage{amssymb}
\usepackage{epsfig}
\usepackage{color}
\usepackage{textcomp}
\usepackage{hyperref}
\usepackage{titlesec}
\usepackage{slashed}
\usepackage{caption}
\usepackage{subcaption}
\def\ee{\end{equation}}
\def\ba{\begin{eqnarray}}
\def\ea{\end{eqnarray}}

\def\bdm{\begin{displaymath}}
\def\edm{\end{displaymath}}

\def\bq{\begin{quote}}
\def\eq{\end{quote}}

 at 10truept

\usepackage{amsmath}
\numberwithin{equation}{section}
\def\ee{\end{equation}}
\def\ba{\begin{eqnarray}}
\def\ea{\end{eqnarray}}

\def\bq{\begin{quote}}
\def\eq{\end{quote}}

 at 10truept

\newcommand{\beq}{\begin{equation}}
\newcommand{\eeq}{\end{equation}}
\newcommand{\beqa}{\begin{eqnarray}}
\newcommand{\eeqa}{\end{eqnarray}}
\newcommand{\bea}{\begin{eqnarray}}
\newcommand{\eea}{\end{eqnarray}}

\def\lesssim{~\mbox{\raisebox{-.6ex}{$\stackrel{<}{\sim}$}}~}

\def\ltap{\ \raise.3ex\hbox{$<$\kern-.75em\lower1ex\hbox{$\sim$}}\ }
\def\gtap{\ \raise.3ex\hbox{$>$\kern-.75em\lower1ex\hbox{$\sim$}}\ }
\def\gl{\ \raise.5ex\hbox{$>$}\kern-.8em\lower.5ex\hbox{$<$}\ }
\def\roughly#1{\raise.3ex\hbox{$#1$\kern-.75em\lower1ex\hbox{$\sim$}}}

\newcommand{\bi}{\begin{itemize}}
\newcommand{\ei}{\end{itemize}}

\def\ltap{\ \raise.3ex\hbox{$<$\kern-.75em\lower1ex\hbox{$\sim$}}\ }
\def\gtap{\ \raise.3ex\hbox{$>$\kern-.75em\lower1ex\hbox{$\sim$}}\ }
\def\gl{\ \raise.5ex\hbox{$>$}\kern-.8em\lower.5ex\hbox{$<$}\ }
\def\roughly#1{\raise.3ex\hbox{$#1$\kern-.75em\lower1ex\hbox{$\sim$}}}

\newcommand{\nede}[0]{_{_{\rm NEDE}}}

\begin{document}

\title{NANOGrav meets Hot New Early Dark Energy and the origin of neutrino mass}

\author{Juan S.~Cruz}
\email{jcr@sdu.dk}
\affiliation{Universe Origins Group and CP$^3$-Origins\\ University of Southern Denmark, Campusvej 55, 5230 Odense M, Denmark}
\author{Florian Niedermann}
\email{florian.niedermann@su.se}
\affiliation{Nordita, KTH Royal Institute of Technology and Stockholm University\\
Hannes Alfv\'ens v\"ag 12, SE-106 91 Stockholm, Sweden}
\author{Martin S.~Sloth}
\email{sloth@sdu.dk}
\affiliation{Universe Origins Group and CP$^3$-Origins\\ University of Southern Denmark, Campusvej 55, 5230 Odense M, Denmark}

\begin{abstract}

It has recently been speculated that the NANOGrav observations point towards a first-order phase transition in the dark sector at the GeV scale~\cite{NANOGrav:2023hvm}. Here, we show that such a phase transition might already have been predicted in the Hot New Early Dark Energy model (Hot NEDE) \cite{Niedermann:2021vgd, Niedermann:2021ijp}. There, it was argued that two dark sector phase transitions are the signature of neutrino mass generation through the inverse seesaw mechanism. In particular, an IR phase transition serves a double purpose by resolving the Hubble tension through an energy injection and generating the Majorana mass entry in the inverse seesaw mixing matrix. This usual NEDE phase transition is then accompanied by a UV counterpart, which generates the heavy Dirac mass entry in the inverse seesaw mass matrix of a right-handed neutrino. Here, we investigate if the UV phase transition of the Hot NEDE model can occur at the GeV scale in view of the recent NANOGrav observations.
\end{abstract}

\maketitle

\newpage


\section{Introduction}

New Early Dark Energy (NEDE~\cite{Niedermann:2019olb,Niedermann:2020dwg,Niedermann:2020qbw,Niedermann:2021ijp,Niedermann:2021vgd,Cruz:2022oqk,Cruz:2023cxy,Cruz:2023lmn} has been established as a leading framework for addressing the Hubble tension through an early-time (pre-recombination) modification of the $\Lambda$CDM model (see \cite{DiValentino:2021izs, Abdalla:2022yfr,Schoneberg:2021qvd} for reviews on cosmological tensions). NEDE is a fast-triggered phase transition in the dark sector at the eV scale. In the Cold NEDE model (see \cite{Niedermann:2019olb, Niedermann:2020dwg, Cruz:2022oqk}), the trigger is an ultra-light axion-like scalar field (ULA) with a mass of order $10^{-27}$ eV, which can simultaneously resolve the $S_8$ tension and the Hubble ($H_0$) tension~\cite{Cruz:2023lmn}. Instead, in the Hot NEDE model, there is no ultra-light scalar field, and the trigger is instead a non-vanishing temperature of the dark sector, with a critical temperature around the eV scale \cite{Niedermann:2021vgd, Niedermann:2021ijp}.

As most Standard Model (SM) particles acquired their mass in the electroweak phase transition (except perhaps the neutrinos), it is natural to speculate that a new eV scale phase transition could be the origin of neutrino masses. In this case, it was previously argued in \cite{Niedermann:2021vgd, Niedermann:2021ijp} that the most natural framework for neutrino mass generation is the inverse seesaw mechanism \cite{Abada:2014vea, Mohapatra:1986bd, Gonzalez-Garcia:1988okv, Deppisch:2004fa}. If we, for simplicity, consider a single generation of neutrinos, then the small meV mass of the active left-handed neutrino, $\nu_L$ is generated through its mass mixing with a right-handed, $\nu_R$, and a sterile neutrino, $\nu_s$. This requires a mass term of the form
\beq
\mathcal{L}_{\nu} = - \frac{1}{2} N^T C M N + \mathrm{h.c.} \,,
\eeq
where $N \equiv (\nu_L, \nu_R^c, \nu_s)^T$, $C$ is the charge conjugation matrix and $M$ is the mass matrix,
\beq
\label{massmatrix}
M = \begin{pmatrix}
	0 & d & 0   \\
	d & 0 & n   \\
	0 & n & m_s
\end{pmatrix}\,.
\eeq
In the inverse seesaw mechanism, it is typically assumed that the high-energy entries, corresponding respectively to a Dirac mixing of $\nu_R$ with $\nu_L$ and $\nu_s$, are in the range $10^{3}$ eV$ \lesssim d \lesssim 10^{11}$ eV and $10^{5}$ eV$ \lesssim n \lesssim 10^{14}$ eV. The Majorana mass, $m_s$,   for $\nu_s$ is a  low-energy entry in the range $\mathrm{eV} < m_s < \mathrm{GeV}$. When diagonalized, this mass mixing matrix gives rise to a light mass eigenstate with mass\footnote{The subscript ``3'' indicates that, for normal ordering, this is the scale of the heaviest of the three mass eigenstates~\cite{Niedermann:2021vgd}.} $m_3 \simeq  m_s  \kappa^2$, with $\kappa = \mathcal{O}(d)/\mathcal{O}(n)  <1$, in addition to a pair of heavy pseudo-Dirac fermions with mass $\mathcal{O}(n)$. It is well-known that when generalized to three generations of neutrinos, the inverse seesaw mechanism can, in this way, explain the observed mass spectrum and mixing pattern~\cite{Abada:2014vea}.

The naturalness of the model is ensured by assigning lepton number $L=1$ to the sterile neutrino, such that the sterile Majorana mass term, $\propto \overline{\nu^c_s} \nu_s$, violates lepton number by two units. It is, therefore, technically natural to have a small, non-vanishing mass~$m_s$, as the lepton symmetry is restored in the limit $m_s \to 0$. In this way, when the mass matrix is diagonalized, also the light active neutrino mass eigenstate $m_3$ is protected from receiving large quantum corrections.

It is a natural possibility, proposed within the Hot NEDE framework \cite{Niedermann:2021ijp, Niedermann:2021vgd}, that the sterile low-energy (IR) Majorana mass term, $m_s$, is generated in the NEDE phase transition with the NEDE boson, $\Psi$, spontaneously breaking lepton symmetry by acquiring a vacuum expectation value (vev). A Yukawa interaction term of the form $g_s \Psi\, \overline{\nu_s^c} \nu_s$ is allowed by assigning lepton number $L=-2$ to $\Psi$.

Of particular interest, in the context of the recent NANOGrav observations~\cite{NANOGrav:2023gor, NANOGrav:2023hfp, NANOGrav:2023hvm} (see also~\cite{Antoniadis:2023utw,Antoniadis:2023zhi,Antoniadis:2023rey}), is the origin of the high-energy (UV) Dirac mixing between $\nu_R$ and $\nu_s$, $10^{5}$ eV$ \lesssim n \lesssim 10^{14}$ eV. In order to generate it dynamically, it is assumed that the dark symmetry group is broken in two steps; first in the UV and then in the IR. Accordingly, the symmetry group of the dark sector has the form
\begin{align}\label{G_D}
G_\mathrm{D} \times G\nede~.
\end{align}
 The breaking of the $G_\mathrm{D}$ in a dark UV phase-transition is then the origin of the UV mass scale $10^{5}$ eV$ \lesssim n \lesssim 10^{14}$ eV, while the IR NEDE phase transition spontaneously breaking $G\nede$ (later identified with a global $\mathrm{U(1)}_L$) is the origin of the IR mass scale $m_s$, within this model. Together they have the potential to solve the Hubble tension as well as explain the origin of the neutrino masses.  It has recently been suggested that the observations of the NANOGrav experiment can be interpreted as evidence of a cosmic phase transition in the dark sector at the GeV scale \cite{NANOGrav:2023hvm, Athron:2023mer, Bringmann:2023opz,Freese:2022qrl}. It is, therefore, natural to ask, at this point, if the UV phase transition leading to the breaking of $G_D$ can be related to the NANOGrav observations.

In \cite{Niedermann:2021ijp, Niedermann:2021vgd}, this symmetry-breaking pattern for $G_\mathrm{D} \times G\nede$ is realized in detail in a Dark Electro-Weak (DEW) model. However, the UV symmetry breaking scale, in its original version, lacked any guidance from data and was assumed to be above the TeV scale. The present study investigates whether the UV symmetry-breaking scale can be lowered to the GeV scale and connected to the proposed interpretation of the NANOGrav observations in terms of a first-order cosmological phase transition.

\section{Hot NEDE and neutrino mass generation}
\label{neutrino_masses}
To generate the $d$, $n$, and $m_s$ mass entries in Eq.~\eqref{massmatrix} through the consecutive symmetry breaking of $G_\mathrm{D} \times G\nede$ and the electroweak symmetry breaking, we need to assume that charges are assigned to  allow for the Yukawa couplings
\begin{align}
 \mathcal{L}_\mathrm{Y} &=-g_{\Phi} \Phi  \overline{\nu_R} \nu_s  - \frac{ g_s}{\sqrt{2}} \Psi \overline{\nu_s^c} \nu_s  + g_H \overline{\nu_R} L^T \epsilon H + \mathrm{h.c.}~, \label{Yukawas}
\end{align}
where $L=(\nu_L, e_L)$ denotes the SM lepton doublet and $H$ denotes the Higgs doublet. The dark sector Higgs field, $\Phi$, is introduced to induce the first UV dark sector breaking of $G_\mathrm{D}$, which generates the Dirac entry $n = g_\Phi v_\Phi / \sqrt{2}$ when $\Phi \to v_\Phi/\sqrt{2}$. Subsequently, the electroweak breaking leads to $d = g_H v_H/\sqrt{2}$ with $v_H = 246\, \mathrm{GeV}$, coupling $\nu_R$ and $\nu_L$ . Finally, as $G\nede$ is broken at the eV scale,  $\Psi \to v_\Psi / \sqrt{2}$, generating the Majorana mass $m_s = g_s v_\Psi $.

The main motivation of this model lies in the fact that we can relate $m_s$ to the cosmological Hot NEDE parameters needed to resolve the Hubble tension. Numerical simulations show that the fraction of NEDE, $f\nede = \Delta V/\rho_{tot}(t_*)$, at the time of the IR phase transition, needed to resolve the Hubble tension, is $f\nede\sim 0.1$. Here $\Delta V$ is the latent heat of the phase transition, and $\rho_{tot}(t_*)$ is the total energy density at the time of the phase transition, $t_*$, corresponding to a redshift $z_*$.

Using that $v_\Psi \simeq g\nede T_d^* / (2 \sqrt{\lambda})$, where $g\nede$ is the dark sector gauge coupling, $\lambda$ is the self-coupling of the NEDE-boson, and $T_*$ is the critical temperature of the NEDE phase transition, one can show that \cite{Niedermann:2021ijp, Niedermann:2021vgd}
\bea\label{mass_sterile}
m_s \approx (1.2 \, \mathrm{eV}) &\times &  \frac{1}{\gamma^{1/4}} \frac{g_s}{g\nede}
\times   \left[\frac{f\nede /(1-f\nede )}{0.1}\right]^{1/4} \left[ \frac{1+z_*}{5000}\right]\,
\eea
where $\gamma= 4\pi \lambda/g\nede^4 \lesssim 1$ corresponds to the super-cooling regime. Super-cooling is the preferred sceneario for solving the Hubble tension in the NEDE framework because it allows for a sizeable fraction of early dark energy. We see that for  $ \gamma g\nede^4 =4 \pi \lambda < g_s^4$ the sterile mass is super-eV as required by the inverse seesaw, at the same time as satisfying the bounds on the cosmological parameters $f\nede$ and $z_*$ to have a successful solution the Hubble tension.

The active neutrinos feel the presence of $\Psi$ through the sterile-active mixing, which mediates a ``secret'' neutrino self-interaction, implying that the free-streaming properties of neutrinos might be affected. It has been argued that the effect on their free-streaming properties at late times leads to the upper bound~\cite{Archidiacono:2013dua,Niedermann:2021vgd} $g_s < 10^{-7} / \kappa^2$ (with $\kappa^2$ accounting for the mixing suppression), which can be shown to lead to an upper bound on the heaviest active neutrino mass ~\cite{Niedermann:2021ijp, Niedermann:2021vgd}
\bea\label{bound_m3}
m_3  <  (0.087\, \mathrm{eV} ) &\times &  c_0^{1/4} \left(\frac{0.4}{\xi_*} \right)^{1/4}\mathrm{e}^{2.1- \frac{1}{\xi^2_*} \sqrt{\frac{f\nede}{1-f\nede}}} \left[\frac{f\nede /(1-f\nede )}{0.1}\right]^{1/4} \left[ \frac{1+z_*}{5000}\right]^{3/4}\,.
\eea
where $c_0$ is a model-dependent factor of order unity and $\xi_*$ is the ratio between the dark and the SM temperature at the time of the phase transition, $\xi_* = T_d/T$. This makes the model highly predictive, also regarding neutrino physics.

Using $m_3 > 0.06 \, \mathrm{eV}$ from oscillation data and imposing an upper bound  on $\Delta N_\mathrm{eff}$ to avoid problems with BBN and CMB, one obtains the narrow allowed range
\begin{align}\label{range_xi}
0.3 < \xi_* < 0.5\,.
\end{align}

\section{NANOGrav and the dark UV phase transition}

If the Hubble tension is a signature of how the neutrinos obtained their mass in the Hot NEDE model, we are suggesting a very predictive framework for the physics of the dark sector. Of particular relevance for the discussion of NANOGrav, the Hot NEDE model predicts two dark sector phase transitions, one close to the eV scale, related to the spontaneous breaking of $G\nede$ and one between $0.1$ MeV and $100$ TeV related to the spontaneous breaking of $G_D$.

NANOGrav has detected a stochastic background of gravitational waves compatible with a first-order phase transition~\cite{NANOGrav:2023hfp,NANOGrav:2023hvm} (see also~\cite{Antoniadis:2023utw,Antoniadis:2023zhi,Antoniadis:2023rey}). The gravitational wave spectrum produced in a first-order phase transition is controlled by the visible sector temperature, $T_*$, at which the phase transition takes place, the inverse duration of the phase transition $\beta$, and its strength,
\begin{align}
\alpha_*= \frac{\Delta V}{\rho_\mathrm{rad}  }\,,
\end{align}
where $\Delta V$ is the latent heat released in the phase transition and $\rho_\mathrm{rad} $ is the total energy density in the radiation fluid comprising contributions from both the visible and dark sector, i.e.\ $\rho_\mathrm{rad}=\rho_\mathrm{rad,vis}+\rho_\mathrm{rad,d}$.
Moreover, the ratio $(8 \pi)^{1/3} H_*/\beta$ determines the typical bubble separation in units of the Hubble radius at the time of the transition, $1/H_*$. The dominant gravitational wave signal can either arise due to the expanding and colliding bubbles or by sound waves produced by the friction between the expanding bubble walls and the radiation plasma. As noted in~\cite{NANOGrav:2023hvm}, the evidence for a signal dominated by vacuum bubbles as the source is stronger than that for sound waves, showing some preference for a \textit{strong} first-order transition. Moreover, such a first-order scenario provides an even better fit to the data than their base model of supermassive black hole binaries by a Bayes factor of 20. Here, we will therefore focus on the case where vacuum bubbles are the main source of the signal.

The gravitational wave spectrum generated would be detected today as (\cite{NANOGrav:2023hvm} and for reviews see~\cite{Caprini:2018mtu,Athron:2023xlk})
\bea
\Omega_{GW}(f)= 1.67\times 10^{-5} A_{b} \left(\frac{\alpha_*}{1+\alpha_*}\right)^2 \left((8\pi)^{1/3}v_{b}H_*\beta^{-1}\right)^{\sigma_{b}} S_{b}(f)\,,
\eea
where the amplitude estimated in the envelope approximation gives $A_b = 0.0049$, the dependence on the nucleation rate is $\sigma_b=2$, and the velocity of the bubble walls is $v_b =1$ (corresponding to a run-away behavior where the bubble walls approach the speed of light in absence of strong friction effects).
The spectral shape function, $S_{b}(f)$, peaks at the frequencies
 \bea
 f_{b} \simeq 48.5 ~\textrm{nHz}\, g_*(T_*)^{1/2}\left[\frac{g_{*s}(T_\mathrm{eq})}{g_{*s}(T_*)}\right]^{1/3}\left(\frac{T_*}{\mathrm{GeV}}\right)\frac{f^*_{b}}{H_*}\,.
 \eea
 with $f_b^* = 0.58/( (8\pi)^{1/3}v_{b}\beta^{-1})$ and $g_*$ and $g_{*s}$ are the effective number of relativistic and entropy carrying degrees of freedom~\footnote{As argued in \cite{Bringmann:2023opz}, we can identify them with their visible sector values  $g_\mathrm{*,vis}(T_*)$ and  $g_{*s,\mathrm{vis}}(T_*)  $ because the dark sector fluid is assumed to be very subdominant.}.
From their observations using pulsar timing arrays, NANOGrav obtains the following parameter bounds (at the $95\%$ confidence limit)~\cite{NANOGrav:2023hvm}
\begin{align}\label{nanograv}
\alpha_* > 0.29, \quad H_*/\beta > 0.038, \quad 0.023 \mathrm{GeV} < T_* < 1.75 \mathrm{GeV}\,.
\end{align}
As mentioned before, the idea is to identify the phase transition potentially observed by NANOGrav with the UV phase transition occurring in hot NEDE. Introducing the maximal fraction of early dark energy that decays in the UV phase transition as $f\nede^\mathrm{UV} = \left[ V_\mathrm{false}(T_d^*) - V_{\mathrm{true}}(T_d=0) \right]/\rho_\mathrm{tot}(T_*)$, we infer the relation\footnote{This assumes that the symmetry broken minimum after the phase transition is close to its vacuum value at $T_d=0$, which is true in a super-cooled phase transition, i.e., $\Delta V = \left[ V_\mathrm{false}(T_d^*) - V_{\mathrm{true}}(T_d=0) \right]$.}
\begin{align}
f\nede^\mathrm{UV} = \alpha_*/1+\alpha_*\,.
\end{align}
To get an idea about the character of the phase transition, it is useful to consider the quantity $\alpha_d = \Delta V / \mathrm{\rho_{\mathrm{rad},d}}$, which compares the released vacuum energy with the dark sector plasma (the visible sector plasma does not couple to the tunneling field). Denoting the dark sector relativistic degrees of freedom as $g_{*,d}$, we have $g_{*,d}(T_*) \rho_\mathrm{rad, vis}^* =  g_{*,\mathrm{vis}}(T_*) \rho_{\mathrm{rad},d}^* / \xi_*^4$, from which we derive
\begin{align}
\alpha_d = \alpha_* \left(1+ \frac{ g_{*,\mathrm{vis}}(T_*)}{ g_{*,d}(T_*) \xi_*^{4}}\right)\,.
\end{align}
Since for the effective number of relativistic degrees of freedom in the visible and dark sector, $g_{*,\mathrm{vis}}(T_*) \gg g_{*,d}(T_*) $, and $\xi_* < 0.5 $ according to \eqref{range_xi}, we conclude that $\alpha_d \gg 17 \alpha_* > 5$. This implies that the change of vacuum energy dominates over the dark sector radiation plasma, which is the signature of a strong super-cooled phase transition.

Another important constraint to consider can be imposed on the effective number of relativistic degrees of freedom from Big Bang Nucleosynthesis (BBN) and CMB observation. If one assumes that all the latent heat is quickly converted into radiation, it amounts to a change in the effective number of relativistic degrees of freedom,
\begin{align}
\Delta N_\mathrm{eff} &= \frac{4}{7}\left(\frac{11}{4}\right)^{4/3} g_{*,d}(T_\mathrm{BBN}) \,\xi^4_\mathrm{BBN}  \nonumber \\
&\simeq 2.2 \, \alpha_* \frac{g_{*s,\mathrm{vis}}(T_\mathrm{BBN})^{4/3}}{g_{*s,\mathrm{vis}}(T_*)^{4/3}}  \frac{g_{*s,d}(T_*)^{4/3}}{g_{*s,d}(T_\mathrm{BBN})^{4/3}} \frac{g_{*,d}(T_\mathrm{BBN})}{g_{*,d}(T_*)} g_{*,\mathrm{vis}}(T_*)
\end{align}
where we used that $\xi^4_* g_{*,d}(T_*) \approx \alpha_* g_{*,\mathrm{vis}}(T_*)$ for $\alpha_d \gg 1$ (in that case $\Delta V \approx \rho_\mathrm{rad,d} $). This also reproduces the result in~\cite{Bringmann:2023opz} when we assume $g_{*,d}$ and $g_{*s,d}$ to be approximately constant.  We can substitute $g_{*s,\mathrm{vis}}(T_\mathrm{BBN}) = 7.4$, $g_{*,\mathrm{vis}}(T_\mathrm{BBN}) = 6.8$ and $g_{*s,\mathrm{vis}}(T_*)  = g_{*,\mathrm{vis}}(T_*) = 86.25$ (assuming the UV phase transition to occur at the GeV scale). This amounts to
\begin{align}
\Delta N_\mathrm{eff} > 7.2 \alpha_* > 2.1\,,
\end{align}
where we substituted the bound on $\alpha_*$ in \eqref{nanograv}. This is clearly in tension with current constraints from BBN and CMB observations, giving rise to the bounds $\Delta N_\mathrm{eff} < 0.39$~\cite{Fields:2019pfx} and $\Delta N_\mathrm{eff} < 0.3$~\cite{Planck:2018vyg}, respectively. This is a common challenge when dark sector phase transitions are invoked to explain the NANOGrav signal~\cite{Bringmann:2023opz}.

Here we offer a possible way out which, in addition, could verify the prediction about the dark sector temperature in~\eqref{range_xi}. If we assume that the dark sector is in thermal contact with the visible sector until after the latent heat is converted into radiation, we have $\xi_*=1$. If both sectors then decouple, $\xi$ starts to decrease as more and more particles in the visible sector fluid become non-relativistic and dump their energy in the fluid when they annihilate. Assuming again that  $g_d$ and $g_{d,s}$ are approximately constant, we find
\begin{align}
 \xi_\mathrm{BBN} = \frac{g_{*s,\mathrm{vis}}(T_\mathrm{BBN})^{1/3}}{g_{*s,\mathrm{vis}}(T_*)^{1/3}}  \xi_* \simeq 0.44 \, \xi_*\,,
\end{align}
which falls safely in the range predicted in the Hot NEDE model in \eqref{range_xi}. This then translates to $\Delta N_\mathrm{eff} \simeq 0.08 \, g_d(T_\mathrm{BBN})$, which is indeed compatible with the observational bound for $g_d(T_\mathrm{BBN}) \lesssim 5$.\footnote{We also mention another possibility to avoid the bound on $\Delta N_\mathrm{eff}$: Instead of assuming a quick conversion of the vacuum bubble after the phase transition into radiation, there is the possibility that the small-scale condensate acts like a fluid with equation of state $w>1/3$ when it is dominated by kinetic (rather than potential or gradient) energy~\cite{Cruz:2023lmn}.  } Whether the thermalization and late decoupling between both sectors, possibly via a neutrino portal coupling, can be achieved in a way that does not conflict with other particle physics constraints remains to be seen.

The finite temperature effective potential for the modulus of a complex scalar $\phi = |\Phi|$, controlling the phase transition, is often written as \cite{Dine:1992wr}
\bea\label{VT1}
V(\phi,T_d) = D(T_d^2-T_0^2)\phi^2 - E T_d\phi^3 +\frac{\lambda^\mathrm{(UV)}_T}{4}\phi^4~.
\eea
where $\lambda^\mathrm{(UV)}_T = \lambda^\mathrm{(UV)} - \Delta \lambda^\mathrm{(UV)}(T_d)$, with $\lambda^{(UV)}$ the field's quartic self-coupling, includes a weak logarithmic temperature dependence. For a generic gauge theory with gauge coupling $g_d$, the parameters can be approximated as $D \sim g_d^2$,   $E \sim g_d^3$, and $T^2_0 \sim (\mu^\mathrm{(UV)})^2 / g_d^2$, where $\mu^\mathrm{(UV)}$ is the field's tachyonic vacuum mass parameter and $\lambda^\mathrm{(UV)} \ll g_d^2$ was assumed.
However, this approximation is not valid in the limit of a strong super-cooled phase transition with $\alpha_d \gg 1$ and a small dark sector percolation temperature $T_* \ll g_d v_\phi = g_d \mu^\mathrm{(UV)} / \sqrt{\lambda^\mathrm{(UV)}}  $. It can be shown that the super-cooled regime can be distinguished from the regime where the above approximation is valid by introducing the parameter~\cite{Niedermann:2021vgd}
\bea\label{def:gamma}
\gamma =\frac{\lambda^\mathrm{(UV)}}{(4\pi E^4)^{1/3}}.
\eea
The effective potential is only ever valid in the regime of\footnote{For improved expressions in the (complementary) high-temperature limit, see~\cite{Arnold:1992rz,Croon:2020cgk,Schicho:2022wty}. } $\gamma \gg 1$, while the super-cooled regime is obtained when $\gamma \lesssim 1$.  In the latter regime, the effective potential instead takes the form (neglecting subdominant scalar contributions)
\bea\label{VT2}
V(\phi,T_d) = V_\mathrm{vac}(\phi) +3T_d^4 \sum_i K(m_i(\phi) /T_d)e^{-m_i(\phi) / T_d} + \ldots\, .
\eea
with $m_i(\phi) \approx g_d v_\phi$ the gauge boson masses and $K(x)$ being a slowly varying function in the range $ 0.1 < |K(x)|< 10$ for $0<x\equiv m_i(\phi)/T_d <20$.  $V_\mathrm{vac}(\phi)$ is the vacuum potential, for concreteness $V_\mathrm{vac}(\phi) = -\mu^2 \phi^2/2 + \lambda \phi^4 / 4  $. The ellipses stand for loop corrections to the vacuum potential. They can be shown to effectively renormalize $\mu^\mathrm{(UV)}$ and $\lambda^\mathrm{(UV)}$ and introduce a weak logarithmic field dependence~\cite{Coleman:1973jx,Kirzhnits:1976ts}. In the more concrete model discussed in the next section, we will therefore identify $\mu$ and $\lambda$ with their one-loop corrected expressions. In this regime, we have~\cite{Niedermann:2021vgd} (defining $\gamma = 4 \pi \lambda^\mathrm{(UV)} / g_d^4$ in accordance with \eqref{def:gamma})
\bea
f\nede^\mathrm{(UV)}\simeq \frac{\pi}{16}\frac{1}{\gamma}\frac{T_d^*}{\rho_\mathrm{tot}(t_*)}\,,
\eea
and
\begin{align}
\alpha_d \approx \frac{15}{8 \pi} \frac{1}{g_{*d}} \frac{1}{\gamma}\,.
\end{align}

In particular, we see that $\alpha_d \gg 1$, as required by the NANOGrav result, can indeed be achieved for $\gamma \ll 1$. This, in turn, implies a sizable fraction of released vacuum energy $f\nede^\mathrm{(UV)}$.
In the super-cooled regime, the duration of the phase transition, $H_*/ \beta$, reads~\cite{Niedermann:2021vgd}
\begin{align}\label{beta_large_mass}
H_*{\beta}^{-1}  \sim 10^{-2} g_d^2   \quad\quad (\text{for}\,\,\gamma \lesssim 1\,\, \text{and} \,\, g_d \lesssim 0.1 )~,
\end{align}
For small gauge coupling $g_d<0.1$, this is in tension with the NANOGrav constraint in  \eqref{nanograv}. On the other hand, for an order unity gauge coupling a more careful analysis, which takes into account order unity factors, will be required for us to decide if the phenomenological bound can be fulfilled. In any case, realizing a slow phase transition with $H_*/\beta \gtrsim 0.01$  within the super-cooled regime (characterized by a sizable energy difference $\Delta V$) constitutes a general model building challenge.

The preferred value of the temperature of the phase transition overlaps with the required value for the UV phase transition in the Hot NEDE model as an explanation of neutrino mass via the inverse seesaw mechanism, specifically $10^{5}$ eV$ \lesssim n \lesssim 10^{14}$ eV. It is, therefore, natural to ask whether the NANOGrav data could be evidence of the UV cosmological phase transition leading to the spontaneous breaking of $G_{\textrm{D}}$ in $G_{\textrm{D}}\times G\nede$. To answer this question, we need to consider a concrete model and check if it allows for the parameters of the super-cooled phase transition required by the NANOGrav observations. This is what we do in the next section.

\section{Dark Electroweak Model}\label{sec:DEW}
For concreteness, we consider the DEW model proposed in \cite{Niedermann:2021vgd}. There it was argued that the minimal choice is to identify $G\nede$ with the global $\mathrm{U(1)}_\mathrm{L}$ lepton number symmetry as the Majorana mass breaks Lepton number, and have $G_\mathrm{D}=\mathrm{SU(2)_D}\times \mathrm{U(1)_{Y_\mathrm{D}}}$, a dark copy of the electroweak group with $Y_\mathrm{D}$ referring to a dark hyper-charge. The gauge couplings associated to the factors of $G_D$ are denoted by $g_d$ and $g'_d$ respectively, and relate as $g_d \simeq g'_d$. It is further assumed that the Higgs $\Phi = (\Phi_+, \Phi_0)^T $ and the sterile $S=(\nu_s, S_-)^T$ transform as doublets with $Y_{\mathrm{D},S}= - Y_{\mathrm{D},\Phi} = -1 $. On the other hand, the NEDE field is promoted to a triplet $\Psi = (\Psi_1,\Psi_2,\Psi_3)^T$ with $Y_{\mathrm{D},\Psi} = 2$, while the SM fields and $\nu_R$ transform as singlets with $Y_{\mathrm{D}} = 0$. In this case, \eqref{Yukawas} becomes
\begin{equation}
 \mathcal{L}_\mathrm{Y} =-g_{\Phi} \overline{\nu_R} S^T \epsilon \Phi  - \frac{ g_s}{2 } \overline{S^c} \epsilon \Delta S + g_H \overline{\nu_R} L^T \epsilon H + \mathrm{h.c.}~, \label{Yukawas2}
\end{equation}
with $\Delta = \Psi \cdot \tau $, $\tau = (\tau_1,\tau_2 ,\tau_3)$ the Pauli matrices and $\epsilon = i \tau_2$. The Higgs potential looks similar to  a dark sector version of the Gelmini-Roncadelli model~\cite{Gelmini:1980re}
\begin{multline}
\label{fullpotential}
V= -\left(\mu^\mathrm{(UV)}\right)^2 \Phi^\dagger \Phi + \lambda^\mathrm{(UV)} \left( \Phi^\dagger \Phi \right)^2  - \frac{ \left(\mu^\mathrm{(IR)}\right)^2}{2} \mathrm{Tr} \left( \Delta^\dagger \Delta\right) + \frac{\lambda^\mathrm{(IR)}}{4} \left[ \mathrm{Tr} \left( \Delta^{\dagger}\Delta \right) \right]^2\\
 + \frac{e-h}{2} \Phi^\dagger \Phi  \mathrm{Tr} \left( \Delta^{\dagger}\Delta \right) + h \Phi^\dagger \Delta^\dagger \Delta \Phi +\frac{f}{4} \mathrm{Tr}\left( \Delta^\dagger \Delta^\dagger\right)   \mathrm{Tr}\left( \Delta \Delta\right) \,.
\end{multline}

The high-energy (UV) breaking is triggered when $\Phi $ picks up its vev, $\Phi \to (0, v_\Phi/\sqrt{2})^T $, breaking $G_\mathrm{D} \to \mathrm{U(1)}_\mathrm{DEM}$. The components of $\Psi$ can be decomposed into neutral, as well as single-charged and double-charged states $\Psi_{\smash{{}^1_2}} =  \pm 1/\sqrt{2} (\Psi_0 \pm \Psi_{++})$ and $\Psi_3 = \Psi_+$. To preserve the dark electromagnetic group, $\mathrm{U(1)}_\mathrm{DEM}$, after symmetry breaking, the NEDE scalar is identified with the neutral component $\Psi_0$.

During the low-energy (IR) NEDE phase transition, the NEDE-boson, $\psi$, acquires a vev $\psi \equiv \sqrt{2} |\Psi_0|\to v_\Psi$, breaking the global $G\nede =\mathrm{U(1)}_\mathrm{L}$ by two units, which, at the same time, gives mass to the active neutrinos.
As a necessary condition for the dark Higgs bosons to be able to acquire the vevs in the phase transitions, we can derive the vacuum  condition from \eqref{fullpotential}
\begin{subequations}
\label{vev_system}
\begin{align}
-\left(\mu^\mathrm{(UV)}\right)^2 + \lambda^\mathrm{(UV)} v_\Phi^2  + \frac{1}{2} \left( e-h \right) v_\Psi^2 &= 0 \,,\\
-\left(\mu^\mathrm{(IR)}\right)^2 + \lambda^\mathrm{(IR)} v_\Psi^2  + \frac{1}{2} \left( e-h \right) v_\Phi^2  &= 0 \,.
\end{align}
\end{subequations}
In order for the inverse seesaw mechanism to be successful, without having to introduce new hierarchies of Yukawa couplings in the dark sector, it is further assumed that $v_\Psi \ll v_\Phi$.
Without fine-tuning, this implies that the two equations in \eqref{vev_system} decouple, provided $e,h \lesssim \lambda^\mathrm{(IR)} v_\Psi^2 / v_\Phi^2 \ll 1$. Since the dark $W'$ and $Z'$ gauge bosons acquire mass in the first UV phase transition, they can potentially spoil these conditions for the IR phase transition to be successful, making it technically unnatural. However, the above condition is technically natural if the dark gauge coupling is sufficiently small, satisfying the conditions $g_d^2 \lesssim \mu^\mathrm{(IR)} / v_\Phi $ and $g_d^4 \lesssim \lambda^\mathrm{(IR)} $ \cite{Niedermann:2021vgd}. The condition $H_*/\beta \approx 0.01\times g_d^2$ for $g_d\lesssim 0.1$ and $\gamma\lesssim 1$  [see \eqref{beta_large_mass}], implies that the NANOGrav observations  $H_*/\beta \gtrsim 0.04$ forces us into a parameter regime with relatively large gauge couplings, $0.1 \lesssim g_d\lesssim 1$, where the lightness of the IR NEDE Higgs boson mass becomes technically unnatural (similar to the naturalness problem with the SM Higgs mass), or allow for new hierarchies of Yukawa couplings in the dark sector.  Avoiding such hierarchies and addressing this naturalness problem is beyond the scope of this work and would require an extension of the model discussed here, where we, for simplicity, assume the decoupling of the two sectors.

In this setup, where $\Phi$ and $\Delta$ decouple in the potential, the effective potential for the UV phase transition is simply given by the dark electroweak theory. The one-loop corrected zero temperature potential can be written in the form
\bea
V_0 =  -\frac{\left(\mu^\mathrm{(UV)}\right)^2}{2}  \phi^2 + \frac{\lambda^\mathrm{(UV)}}{4} \phi^4 + 2B v_\Phi^2 \phi^2-\frac{3}{2} B \phi^4 + B \phi^4\ln\left(\frac{\phi^2}{v_\Psi^2}\right)
\eea
with $\phi = \sqrt{2} |\Phi|$ being the modulus of the complex $\Phi$ field. In this case, the terms in the finite temperature potential controlling the UV phase transition in the high-temperature regime are given by \cite{Dine:1992wr} 
\bea
\label{eq:DET0}
D ~&=& \frac{1}{8 v_\Phi^2}(2 m_{W'}^2 + m_{Z'}^2)\nonumber\\
E ~&=& \frac{1}{4 \pi v_\Phi^3}(2 m_{W'}^3 + m_{Z'}^3)\nonumber\\
T_0^2&=&\frac{\left(\mu^\mathrm{(UV)}\right)^2}{2D}-\Delta T_0^2 \nonumber\\
\eea
with $m_{W'}^2 = m_{Z'}^2 = g_d^2 v_\phi^2$ and
\bea
\Delta T_0^2 = \frac{2}{D} B v_\Phi^2,\qquad \textrm{and} \qquad B= \frac{3}{64 \pi^2 v_\Phi^4}\left(2 m_{W'}^4+m_{Z'}^4\right).
\eea

From the equations above, it is evident that $T_0 \to 0$, when the effective one-loop zero-temperature mass vanishes for $\left(\mu^\mathrm{(UV)}\right)^2 = 4 B v_\Phi^2$. Using $\lambda^\mathrm{(UV)}=\left(\mu^\mathrm{(UV)}\right)^2 / v_\Phi^2$ that implies a lower bound on the tree-level self-coupling $\lambda^\mathrm{(UV)} > 4 B = 36 g_d^4/(64 \pi^2)$. Before reaching that lower bound, when (neglecting the log term in the zero temperature potential)  $\lambda^\mathrm{(UV)} \to 6 B =   54 g_d^4/(64 \pi^2) $ from above, we have that the effective one-loop interaction $\lambda^\mathrm{(UV)}_\textrm{int}= \lambda^\mathrm{(UV)}- 6B$ goes to zero
\bea
\lambda^\mathrm{(UV)}_\textrm{int} \to 0 \qquad \text{when} \qquad \lambda^\mathrm{(UV)} \to  6 B = \frac{54 g_d^4}{64 \pi^2},
\eea
which is also the limit of extreme super-cooling where
\bea
\gamma = \frac{\lambda^\mathrm{(UV)}_\textrm{int}}{(4\pi E^4)^{1/3}} \to 0 \qquad \text{when} \qquad \lambda^\mathrm{(UV)} \to  6 B .
\eea
Within the super-cooled regime, $\gamma\lesssim 1$, we then have \cite{Niedermann:2021vgd}
\bea
\alpha_{d} \approx \frac{15}{8\pi} \frac{1}{g_\textrm{rel,d}}\frac{1}{\gamma},
\eea
where $\alpha_d \to \infty$ as $\gamma \to 0$.

We showed earlier, Eq.~\eqref{range_xi}, that the model requires $0.3 < \xi_* <0.5$. In \cite{Niedermann:2021vgd} it was found that for $\gamma \lesssim 1$, we have
\bea\label{xi}
\xi_* \simeq (6\,\gamma\,\alpha_* )^{1/4}
\eea
Hence, for $\alpha_*\approx 0.3$, as required by NANOGrav, and  $\xi_* \approx 0.44$, we obtain $\gamma=0.02$, which corresponds in turn to $\alpha_d \approx 30/g_{\textrm{rel,d}}$. We can also estimate the energy scale of the phase transition. In the supercooling-regime where $\lambda^\mathrm{(UV)} \approx 6B$, we infer from \eqref{eq:DET0} that $T_0 \approx \sqrt{\Delta T^2_0} \sim g_d v_\phi$. Requiring $T_0\approx  $ GeV, we therefore need $v_\Phi \approx  $ GeV for $g_d \approx 0.5$.

\section{Discussion}

In this short communication, we have investigated whether a first-order cosmological phase transition (FOPT) at the GeV energy scale predicted within the Hot NEDE model~\cite{Niedermann:2021vgd, Niedermann:2021ijp} is consistent with the recent observation of a gravitational wave signal made by the NANOGrav Collaboration~\cite{NANOGrav:2023hvm}. A first-order phase transition dominated by bubble wall collisions has been seen to have a Bayes factor of around 20 when compared to an explanation of the signal via supermassive black hole binaries and represents a good incentive to look more in-depth at models that feature this type of transition. While more work and data collection is required to verify (or exclude) whether the NANOGrav observations are indeed indicating the existence of a strong super-cooled phase transition at the GeV energy scale, it is interesting to understand what would be the most likely theoretical explanation for such a phase transition. More precise predictions within a theoretical framework will enable us to test our models more easily.

We have briefly reviewed the Hot NEDE framework, whose main achievement is alleviating the Hubble tension. The model features a phase transition at the eV scale, induced by the NEDE boson field, $\psi$, which is thermally triggered as a new global minimum in the potential opens up when the hidden sector has cooled down sufficiently. It is possible to embed this mechanism in an explicit microphysical description where the NEDE phase transition is a signature of neutrino mass generation in the inverse seesaw mechanism. We explore the Dark Electroweak model (DEW) as a concrete realization that can generate neutrino masses. In this model, the high-energy (UV) DEW symmetry breaking at the GeV scale generates the high-mass neutrino mixing matrix elements required by the inverse seesaw mechanism. On the other hand, the low-energy (IR) spontaneous breaking of the lepton symmetry generates an eV sterile mass,  as also required by the inverse seesaw mechanism, and resolves the Hubble tension by a pre-recombination energy injection. In view of the newly suggested parameters for a FOPT from the fits to the NANOGrav signal,  the DEW model, with its richer dark sector featuring an additional FOPT at the MeV to TeV energy scale, becomes very interesting.

NANOGrav fits suggest a bound for the strength of a bubble-dominated transition of $\alpha_* > 0.29$, a percolation temperature $T_* \in [0.023\, \text{GeV}, 1.75\, \text{GeV}]$ and bubble nucleation rate per Hubble patch at percolation, $\beta H_*^{-1} < 20.9$ (all at $95\%$ C.L.). In the present work, we find that the aforementioned parameters could, indeed, be compatible with the UV phase transition already predicted within the Hot NEDE framework for explaining the origin of neutrino masses.

The main problem with explaining the NANOGrav observations by a dark phase transition comes from the change in the effective number of relativistic degrees of freedom $\Delta N_{\textrm{eff}}$ if all the latent heat of the phase transition is converted into radiation. In this case, BBN bounds on $\Delta N_{\textrm{eff}}$ are a challenge. However, we proposed that if the dark sector is thermalized with the visible sector after the phase transition but decouples before BBN at the MeV scale, then this could lower  $\Delta N_{\textrm{eff}}$ to an acceptable level with $\xi_* = 0.44$ at the time of BBN. This would interestingly also be consistent with the required range, $0.3 <\xi_* < 0.5$, predicted by the Hot NEDE model by different neutrino physics constraints and can thus be viewed as an additional verification of the Hot NEDE predictions. We speculate that this thermalization and subsequent decoupling could happen f.ex. through the neutrino portal in the model.  Another possibility is a stiff fluid component after the phase transition lowering $\Delta N_{\textrm{eff}}$.

In the specific DEW embedding of the scenario of neutrino mass generation via the inverse seesaw mechanism in the Hot NEDE model, we need relatively large dark gauge couplings, $0.1 \lesssim  g_d  \lesssim 1$, similar to the gauge couplings in the SM electroweak theory, in order to achieve a slow enough phase transition in accordance with the NANOGrav observations, $H_*/\beta> 0.04 $. However, this induces a naturalness problem for the lightness of the IR NEDE Higgs boson, unless we allow for new hierarchies of Yukawa couplings in the dark sector. Within the discussed model, the small Yukawa hierarchies and naturalness of the NEDE boson mass would require smaller gauge couplings and hence a faster phase transition. Addressing the Yukawa hierarchies and the naturalness of the NEDE-boson mass while explaining the NANOGrav observations would require a relaxation of the assumption that the UV and IR Higgs potentials decouple, or some extension of the DEW model discussed here.

Although it requires more work to test these possibilities in all details, it is encouraging that our initial study shows that the UV Hot NEDE phase transition can indeed be a strong super-cooled phase transition, with the required strength $\alpha_* > 0.29$ at the GeV scale, and that the NANOGrav observations can be connected to the origin of the neutrino masses and the Hubble tension through the Hot NEDE model.

It will be interesting to investigate further the possibility that the NANOGrav observations, the Hubble tension, and the $S_8$ are all manifestations of the origin of neutrino mass in the Hot NEDE model.

\begin{acknowledgments}
J.S.C. and M.S.S. are supported by Independent Research Fund Denmark grant 0135-00378B. The work of F.N. is supported by VR Starting Grant 2022-03160 of the Swedish Research Council.
\end{acknowledgments}

\bibliography{Hot_NEDE}

\end{document}